\documentclass[showpacs,twocolumn,aps,preprintnumbers,letterpaper]{revtex4}
\usepackage{amsmath,amssymb}
\usepackage{epsfig}
\usepackage{graphicx}
\usepackage{amsmath}
\usepackage{slashed}
\usepackage{amsfonts}
\usepackage{epstopdf}
\usepackage{color}

\newcommand{\be}{\begin{equation}}
\newcommand{\ee}{\end{equation}}
\newcommand{\bear}{\begin{eqnarray}}
\newcommand{\eear}{\end{eqnarray}}
\newcommand{\ba}{\begin{array}}
\newcommand{\ea}{\end{array}}

\def\be{\begin{eqnarray}}
\def\ee{\end{eqnarray}}
\def\bea{\be}
\def\eea{\ee}

\def\roughly#1{\mathrel{\raise.3ex\hbox{$#1$\kern-.75em%
\lower1ex\hbox{$\sim$}}}}

\begin{document}

\title{Disorder in the Sachdev-Yee-Kitaev Model}

\author{Yizhuang Liu}
\email{yizhuang.liu@stonybrook.edu}
\affiliation{Department of Physics and Astronomy, Stony Brook University, Stony Brook, New York 11794-3800, USA}

\author{Maciej A. Nowak}
\email{nowak@th.if.uj.edu.pl}
\affiliation{M. Smoluchowski Institute of Physics and Mark Kac Complex Systems Research Center, Jagiellonian University, S. \L{}ojasiewicza 11, PL 30-348 Krak\'{o}w, Poland }

\author{Ismail Zahed}
\email{ismail.zahed@stonybrook.edu}
\affiliation{Department of Physics and Astronomy, Stony Brook University, Stony Brook, New York 11794--3800, USA}


\date{\today}
\begin{abstract}
We give qualitative arguments for the mesoscopic nature of the Sachdev-Yee-Kitaev (SYK) model in the holographic
regime with $q^2/N\ll 1$ with $N$ Majorana particles coupled by antisymmetric and random interactions  of range $q$. 
Using a stochastic deformation of the SYK model, we show that  its characteristic determinant  obeys
a viscid Burgers equation with a small  spectral viscosity in the opposite regime with $q/N=1/2$, in  leading order.
The stochastic evolution of the SYK model can be mapped onto that of  random matrix theory,
with universal Airy oscillations at the edges.  A spectral hydrodynamical estimate for the relaxation
of the collective modes  is made.
 \end{abstract}


\maketitle

\setcounter{footnote}{0}


\section{Introduction}
The understanding of how entropy is produced and developed in heavy-ion collisions  at ultra-relativistic 
energies is the subject of intensive interest at collider energies~\cite{ED}. Key to this is the concept of thermalization
and the time it takes to reach it~\cite{EDHOLO,RAJU,JANIK,MORE,EDSTRING}. More generally, there is a wide theoretical interest in the understanding of
non-perturbative entropy formation in quantum processes ranging from atomic systems at the unitarity limit~\cite{ATOM,COLD} 
to string theory using black-holes~\cite{STRINGBH}. 

The SYK model consists of $N$ quantum mechanical fermions with Gaussian distributed 
random couplings of rank-$q$ and strength $J$.  The model is solvable at large $N$ and fixed
$J$  by a saddle point approximation where a special class of Feynman graphs is selected. 
Originally, this model was proposed by Sachdev and Yee~\cite{SY}
to describe quantum spin fluids.  More recently, Kitaev~\cite{KITAEV} has suggested 
the model to shed light 
on holography, by arguing that the large $N$ limit of the model is dual to a black hole 
in an emergent AdS$_2$ space-time. A number of investigations have since 
followed~\cite{MANY,MALDACENA,POLCHINSKY,VERBA}.

The SYK model offers a simple framework for discussing the formation of black holes in quantum mechanics~\cite{KITAEV}.
In the regime $q^2/N\ll 1$,  numerical analyses~\cite{VERBA,POLCHINSKY}  support the
existence of a chaotic regime described by random matrix theory  at late times. The  chaotic regime signals
the onset of a black hole. The purpose of this paper is to stress  the mesoscopic nature of the SYK model
throughout its ballistic, diffusive and ergodic regimes. In the opposite regime with $q^2/N\gg 1$, we make use 
of spectral determinants,  to show how a viscid fluid description emerges 
with a small spectral viscosity that maps exactly on random matrix theory. This regime is dominated by
planar diagrams in leading order.

The organization of the paper is as follows: in section 2 we 
briefly outline the SYK  model and discuss its bulk spectral
distribution in the holographic  limit with $q^2/N\ll 1$. In section 3, 
we provide a qualitative description of the mesoscopic
nature of the model and give a simple estimate for the ergodic time. 
In section 4, we discuss the opposite limit with $q^2/N\gg 1$ using
the SYK characteristic determinant. We show that it obeys a
viscid Burgers equation which is analogous to the one derived using 
random matrix theory for the GUE ensemble. The inviscid 
equation gives rise to a semi-circular distribution, while the viscid equation
gives rise to Airy universality at the edge of the spectrum. We use it to estimate
the contribution of the edge states to the partition function at low temperature.
We also suggest a spectral hydrodynamical estimate for the stochastic relaxation of the
collective modes. Our conclusions are in section 5.

\section{SYK model}

The SYK model  consists  of $N$-Majorana fermions with $q$-interactions in $0+1$-dimensions, 
with random and antisymmetric couplings. The corresponding quantum-mechanical Hamiltonian say for $q=4$ 
is 

\be
\label{1}
H=\sum_{a<b<c<d}J_{abcd}\,\psi^a\psi^b\psi^c\psi^d
\ee
with $N$-Majorana fermions $\psi_a$  of flavor $a=1,...,N$. The couplings $J$ are antisymmetric in all entries 
and randomly sampled from a Gaussian distribution with zero mean but fixed variance $N_q=(q-1)!J^2/N^{q-1}$. 
All units are set by $J$  which will be set to 1.  As operators in a Hilbert space, the
flavored Majorana $\psi^a$ map onto the representations of the Clifford algebra $Cl[\frac N2]$ as realized by $\gamma^a$ 
matrices

\be
\label{1X}
H\rightarrow \sum_{a<b<c<d}J_{abcd}\,\gamma^a\gamma^b\gamma^c\gamma^d\equiv \sum_AJ_A\Gamma_A
\ee
wich is $L\times L$ valued with $L=2^{\frac N 2}$. (\ref{1X}) refers to a sum of sparse matrices with
$N^4$ random weights. (\ref{1}-\ref{1X}) exhibit particle-hole symmetry which is
enforced by an anti-unitary operation. As a result the spectra exhibit some degeneracy for some values of
$N$ modulo 8 (Bott periodicity)~\cite{POLCHINSKY,VERBA}.


The Hamiltonian in~(\ref{1X})  is bounded and symmetric.
The edge states map onto the low-lying
$N$-body  excitations close to the ground state, and the central states 
map onto the high lying $N$-body excitations. The  latters 
follow a Gaussian distribution~\cite{POLCHINSKY,VERBA}. 
Specifically, consider  the average  partition function 

\be
\label{PARTX}
\left<{\mathbb Z}(\beta)\right>_J\equiv \left<{\rm Tr}\left(e^{-\beta H}\right)\right>_J=\sum_k\frac{\beta^{2k}}{(2k)!}\left<H^{2k}\right>_J
\ee
where uniform convergence is assumed (this is likely upset at the edges~\cite{MATH}).
Formally, the moments in  (\ref{PARTX}) are

\be
\label{PARTXX}
\left<H^{2k}\right>_J=\frac{1}{N^k_q}
\sum_{i_1...i_k}\sum_{j_1...j_{2k}}{\rm Tr}\left(\Gamma_{j_1}\Gamma_{j_2}...\Gamma_{j_{2k}}\right)
\ee
with typically $(j_1j_2...j_{2k-1}j_{2k})=(i_1i_1...i_ki_k)$. Since the $\Gamma^\prime$s need at least one common factor in order to anti-commute, and these pairs form only a small fraction of all pairs, we can throw away the anti-commutators in (\ref{PARTXX}) 
at large $N$~\cite{VERBA,MATH}. 
For a fixed sequence $(i_1i_2...i_k)$,
the trace in each term of the $j$ contribution is $L$ and there are  $(2k-1)!!$ such contributions. The final sum over the 
sequences $(i_1i_2...i_k)$ gives $(C_N^q)^k$

\be
\label{MOM}
\left<H^{2k}\right>_J\approx L\, (2k-1)!!\left(\frac{C_N^q}{N_q}\right)^k
\ee
The partition function  at high temperature (small $\beta$)  is

\be
\label{PARTXXX}
\left<{\mathbb Z}(\beta)\right>_J\approx L \,e^{\frac {\beta^2C_N^q}{2N_q}}\approx L\, e^{\frac {N\beta^2}{2q}}
\ee
with the corresponding bulk entropy

\be
S\approx N{\rm ln}\sqrt{2}-\frac N{2qT^2}
\ee
The inverse Laplace transform of (\ref{PARTXXX}) gives a Gaussian distribution of the central
eigenvalues

\bea
\label{GAUSS1}
\rho(E) =&&\int_C d\beta\, e^{\beta E}\left<{\mathbb Z}(\beta)\right>_J\nonumber\\
\approx &&\int_C d\beta L\,e^{\beta E+N\beta^2/q}\approx L\, e^{-\frac {qE^2}{2N}}
\eea
Overall, (\ref{GAUSS1}) is in agreement with the arguments 
and numerics presented in~\cite{MALDACENA,VERBA,POLCHINSKY}. 
Eq.~(\ref{GAUSS1}) fails at the edges of the spectrum~\cite{MALDACENA,POLCHINSKY,VERBA}. 
For $q^2/N\ll 1$ the symmetric edges expand as $\pm N\lambda_0$ with an exponential growth 
of states away from the edges given by ${\rm sinh}\,({2cN|E-N\lambda_0|})^{\frac 12}$
in the triple scaling limit~\cite{POLCHINSKY}.

\section{Mesoscopy}

\begin{figure}[h!]
\begin{center}
 \includegraphics[width=8cm]{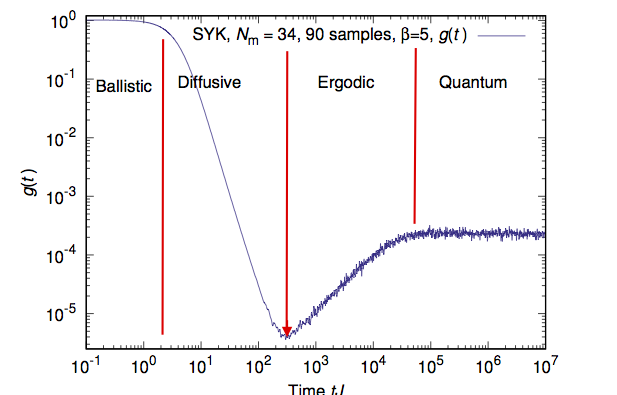}
  \caption{Spectral form factor for the SYK model from~\cite{POLCHINSKY}. We super-imposed on the figure four mesoscopic regimes. An arrow points at the time corresponding to  the ergodic (Thouless) time. }
  \label{fig_FF}
 \end{center}
\end{figure}

In~\cite{SUVRAT,POLCHINSKY}  the deformed spectral form factor was defined as 

\be
\label{2}
g(\beta, t)=\frac{\left<\left|\mathbb Z(\beta+it)\right|^2\right>_J}{\left<\mathbb Z(\beta)\right>_J^2}
\ee
and analyzed both analytically and numerically in~\cite{POLCHINSKY} with a sample of the results shown in Fig.~\ref{fig_FF}. 
The features shown are generic of mesoscopic systems with multi-fermion induced interactions
such as those developed in the disordered QCD vacuum~\cite{CHIRAL} (and references therein). 

A useful formula for discussing mesoscopic systems is the semi-classical form of the 
spectral form  factor at large $\beta$ or small temperatures~\cite{HO}

\be
\label{4}
g(t)\approx \frac{|t|p(t)}{(2\pi)^2}
\ee
for times much smaller than the quantum (Heisenberg) time, $t<t_H=1/\delta$ with $\delta=2\pi/L$
the quantum energy spacing.  Here $p(t)$ is the classical return probability, typically of the form

\be
\label{RET}
p(t)=\left<\left|\left<J|\gamma^a(t)\gamma^a(0)|J\right>\right|^2\right>_J
\ee
The microscopic and quantitative many-body analysis of  (\ref{RET})  will be presented elsewhere~\cite{ELSE}.
Qualitatively, (\ref{RET}) can be thought as the probability of return for
a given flavor undergoing anomalous Brownian motion in a linear volume $V_1=L$. 
Each random walk
spreads in a time $t$ an effective  squared distance $X^2\approx t^{\Delta}$ with $\Delta$ the 
anomalous diffusion exponent ($\Delta=1$ for the canonical Brownian random walk). 
 A simple estimate of the return probability (\ref{RET}) in this random walk approximation is
$p(t)\approx V_1/X$. The ergodic (Thouless) time is reached  when the random walks fill out 
the effective volume $X\approx V_1$ causing $p(t)\approx 1$, that is $t_E\approx L^{\frac 2\Delta}$.

In the   diffusive regime with $t_B<t<t_E$, (\ref{4}) becomes

\be
\label{6}
g(t_B<t<t_E)\approx \frac{|t|}{(2\pi)^2}\,\frac{L}{t^{\frac \Delta 2}}
\ee
For all super-diffusive random walks with 
$\Delta>2$, (\ref{6})  is in qualitative agreement with the slope in Fig.~\ref{fig_FF}. 
In particular, for $\Delta=8$, the result (\ref{6})  is in agreement with the  numerical results and estimates
in~\cite{VERBA,POLCHINSKY}. 
The ballistic time is identified as $t_B\approx L^0$
below which the left plateau is seen in Fig.~\ref{fig_FF}.
In the ergodic regime with $ t_E<t<t_H$ we have $p(t)\approx 1$, and 

\be
g(t_E<t<t_H)\approx  \frac{|t|}{(2\pi)^2}\,
\ee
grows linearly with time in overall agreement with the rise in Fig.~\ref{fig_FF}. 
In the Heisenberg regime with $t>t_H$, the  spectral form factor is dominated by 
the self-correlation for a single energy level which is normalized to a delta-function 
in energy space $g(E\rightarrow 0)= \delta(E)$, and translates to a constant in time

\be
\label{7}
g(t>t_H)=\frac 1{2\pi}
\ee
which is the right plateau  in Fig.~\ref{fig_FF}. 
As in mesoscopic systems, we note the hierarchy of  times

\be
\label{8}
t_B\approx L^0<t_E\approx L^{\frac 2\Delta}<t_H\approx L
\ee
The ergodic regime is universal and follows from random matrix theory and symmetries as observed in~\cite{POLCHINSKY,VERBA}.
 In the presence of time-reversal symmetry the counting of paths in (\ref{4}) is increased by a factor of 2. The particularly
 short  time $t_S\approx {\rm ln}L$ reported  in~\cite{ALTLAND} is of the order of the  Ehrenfest time and may be a signal for the  
 loss of quantum coherence at the edge of  the ballistic regime.

\section{Random matrix limit}

In the $q^2/N\ll 1$ limit, the SYK model provides a quantum mechanical realization of the holographic 
principle as discussed by many~\cite{MALDACENA,POLCHINSKY,VERBA,MANY}.  By increasing the $q$-range
of the random interaction the model undergoes a transition to a random matrix regime a situation similar
to the one encountered in the context of quantum spin glasses~\cite{MATH}. In this section we specialize to
the case with maximum randomness with $q/N=1/2$ which is  opposite to the holographic 
regime. This regime is dual to random matrix theory and chaotic for all time scales in leading order,
as we now show.

\subsection{Ergodic evolution}

To streamline the counting for the case with $q/N=1/2$, it is more convenient to re-define (\ref{2})  
using  the new normalization for the $q$-range couplings 

\be
H=\frac{1}{(C_{N}^{p_N})^{\frac{1}{2}}}\sum_{J_{p_n}}\alpha_{J_{p_n}}\Gamma_{J_{p_n}}
\ee
with 

\bea
&&\Gamma_{J_{p_N}}\equiv \gamma_{i_1}\gamma_{i_2}...\gamma_{i_{p_n}}\nonumber\\
&& J_{p_n}={1\le i_1<i_2<...<i_{p_n}\le N}
\eea
 a typical basis element of  rank-$p_n$ in the minimal representation of the  Clifford algebra $Cl([\frac N2])$. There are  
$C_N^{p_n}$ such basis elements, and they all satisfy $\Gamma_J^2=1$.
The characteristic determinant for the SYK model is defined as

\be
\label{9}
{\bf {\Psi}}[\tau, z]=\left<{\rm det}(z-H)\right>_J=\int d[J]\,{\bf P}(\tau, J)\,{\rm det}(z-H)\nonumber\\
\ee
with the measure ${\bf P}(\tau, \alpha_J)\approx e^{-\frac 1{2\tau} \alpha_{J_i}\alpha_{J_i}}$, 
and ${\bf P}(0,\alpha_J)\approx \delta(\alpha_J)$. 
We  note that the measure
reduces to a delta-function as $\tau\rightarrow 0$, and asymptotes a Gaussian 
as $\tau> 1$ which is the SYK model. 
Eq.~(\ref{9}) provides  a  stochastic 
deformation of the SYK model with vanishing  couplings as $\tau\rightarrow 0$, 
much like in the random matrix deformation  in~\cite{DET}. 

To analyze (\ref{9}) we set $N=2n$ and specialize to the case $p_n=n$.
This is the case with maximum range for the random couplings.
With this in mind, we unwind the determinant using Grassmannians, 
and carry the Gaussian integration over the random couplings $\alpha_J$ to obtain

\be
\label{CDT}
\Psi[\tau, z]=\int D\chi D\bar \chi e^{-z\bar \chi \chi+\frac{\tau}{2C_{2n}^{n}}\sum\bar \chi \Gamma_{n}\chi \bar \chi \Gamma _{n}\bar \chi}
\ee
We now use a Fierz re-arrangement of the 4-Grassmannian induced interaction

\bea
\label{CDT1}
&&\frac{1}{2C_{2n}^{n}}\sum \bar \chi \Gamma_{n}\chi \bar \chi \Gamma _{n}\bar \chi=\nonumber \\
&&-\frac{1}{2L}\bar \chi \chi \bar \chi \chi-\frac{(-1)^{n}}{2L}\bar \chi \Gamma_{2n}\chi \bar \chi \Gamma_{2n}\chi\nonumber \\ 
&&-\frac{1}{2L}\sum_{A\ne{0,2n}}\frac{N_{A}^+-N_{A}^-}{C_{2n}^{n}}\bar\chi \Gamma_{A}\chi \bar\chi \Gamma_{A}\chi
\eea
with 

\bea
&&N_{p}^{+}=(-1)^{p+n}\sum_{k=0}^{[n/2]}C_{p}^{2k}C_{2n-2k}^{n-2k}\nonumber\\
&&N_{p}^{-}=(-1)^{p+n}\sum_{k=0}^{[n-1/2]}C_{p}^{2k+1}C_{2n-2k-1}^{n-2k-1}
\eea
For any large  $n$, we have

\bea
\label{SUPP}
&&N_{2p+1}^+-N_{2p+1}^{-}=0\nonumber\\ 
&&\frac{|N_{2p}^{+}-N_{2p}^{-}|}{C_{2n}^n}\approx \frac{(2p-1)!!}{2^{p}n^{p}}
\eea
As a result the third line contributions to the Fierz re-arrangement in (\ref{CDT1}) are all of order $1/n$ 
in comparison to the first two lines, and subleading.
Therefore (\ref{CDT1}) simplifies to

\be
\label{CDT2}
-\frac{1}{2L}\bar \chi \chi \bar \chi \chi-\frac{(-1)^{n}}{2L}\bar \chi \Gamma_{2n}\chi \bar \chi \Gamma_{2n}\chi
\ee
Since 
\be
\Gamma_{2n}^2=(-1)^\frac{(2n-1)(2n)}{2}=(-1)^{n(2n-1)}=(-1)^n
\ee
$(-1)^{n/2}\Gamma_{2n}$ squares to one and we can write  (\ref{CDT2}) as

\be
\label{CDT3}
-\frac{1}{L}\bar \chi_{+}\chi_+\bar \chi_{+}\chi_+-\frac{1}{L}\bar \chi_{-}\chi_-\bar \chi_{-}\chi_-
\ee
The labels $\pm$ refer to the positive-negative eigenvalues of the chirality matrix  $(-1)^{n/2}\Gamma_{2n}$.
This result is physically expected as the  Fierzing in (\ref{CDT1}) rescinds the 4-Fermi induced interaction
into all spin channels in the large Hilbert space.  In leading $1/n$,  all the spin bearing channels wash out, except
for the scalar and pseudo-scalar channels with each carrying an effective coupling of  $1/2L$. The chiral copies
in (\ref{CDT3}) reflect on the particle-hole symmetry noted in~\cite{LUKAS,SUBIR,POLCHINSKY,VERBA}.

Therefore at large $n$, the characteristic determinant (\ref{CDT}) splits into two chiral copies with

\bea
\label{LN1}
&&\Psi[\tau, z] \approx \Psi_{+}[\tau, z] \Psi_{-}[\tau, z] \nonumber\\
&&\Psi_{\pm}[\tau, z]=\int D\chi_\pm  D\bar \chi_\pm 
e^{-z\bar \chi_\pm \chi_\pm-\nu_L\tau\bar \chi_{\pm}\chi_\pm\bar \chi_{\pm}\chi_\pm}\nonumber\\
\eea
with $\nu_L=1/L$. It follows that
each of the chiral copies in (\ref{LN1}) close under ergodic evolution (reverse diffusion) 

\bea
\label{LN2}
\partial_{\tau}\Psi_{\pm}= -\nu_L \partial_{zz}\Psi_{\pm} \qquad with \qquad
\Psi_{\pm} (0,z)=z^{L/2}
\eea
Using the complex Cole-Hopf transformation for the characteristic determinant
$f_L=\partial_z{\rm ln}\Psi_{\pm}/\tilde L$ with $\tilde L=L/2$, (\ref{LN2}) for the SYK model maps onto the viscid Burgers equation

\be
\partial_\tau f_L+f_L\partial_zf_L=-\nu_L\partial_{zz}f_L
\label{Burgers}
\ee
with $\nu_L$ playing the role of a (negative) spectral viscosity~\cite{PIOTR}. 
In terms of the (cold) entropy $S/N\approx {\rm ln 2}/2$~\cite{MALDACENA},
the spectral viscosity is $\nu_L=1/e^S$.  
We note that (\ref{LN2}-\ref{Burgers})  map onto 
the ergodic equation for the characteristic determinant of the  unitarity ensemble of random 
matrix theory of finite size $L/2$ and $\beta_D=2$~\cite{PIOTR}. This mapping together with the 
semi-circular distribution (see below) 
guarentee that the spectral form factor in (\ref{2}) is also of the general form shown in Fig.~\ref{fig_FF} in
the random matrix regime with $q/N=1/2$ and in leading order.

\subsection{ Airy universality}

We now focus on one of the two chiral copies and study its spectrum. 
The formal solution to (\ref{LN2}) is

\be
\label{LN3}
\Psi_{\pm}[\tau, z]=\left(\frac{1}{4\pi\nu_L\tau}\right)^{\frac 12}\int_{\mathbb C}dz^\prime 
e^{\frac{1}{4\nu_L\tau}(z-z^\prime)^2}{z^{\prime L/2}}
\ee
which is the convolution of the diffusion kernel with the initial condition in $L$-space. 
The $L$-saddle point 
approximation to (\ref{LN3}) yields the  Cole-Hopf transform

\bea
\label{LCH}
f_L[\tau, z]\approx \frac 1{\tau}(z-z_+)=\frac{1}{z_+}
\eea
with $2z_+=z+\sqrt{z^2-4\tau}$. Eq.~(\ref{LCH}) acts as a Coulomb-like potential
for the macroscopic spectral density of eigenvalues with ($\tilde L=L/2$)

\be
\label{DENSITY1}
\rho(\tau, \lambda)=\frac {\tilde L}{\pi}\,{\rm Im}f_L[\tau, z=\lambda]
\approx  \frac {\tilde L}{2\pi\tau}\left(4{\tau} -\lambda^2\right)^{\frac 12}
\ee
which is semi-circular.

Eq.~(\ref{DENSITY1}) can also be shown to follow from the moment analysis of $H$,  if we were to note that 
all crossing diagrams are suppressed by powers of the ratio in (\ref{SUPP}), 
in comparison to the non-crossing diagrams. As a result only the planar contributions are retained 
 for $N=2n$ and $p_n=n$  at large $n$, leading to the standard Pastur equation for the resolvent and 
 a semi-circle.

The key  feature of the semi-circle are its
edges at $\pm \sqrt{4\tau}$ with an accumulation of  states of order $L\lambda^{3/2}$ which 
suggests
the microscopic re-scaling (unfolding) at the origin of the Airy universality 
(soft-edge universality). This follows from either the rescaled expansion around the saddle point in
(\ref{LN3})~\cite{DET}, or the shock analysis of the viscid  Burgers equation~\cite{PIOTR}, with the result

\be
\label{DETX}
\Psi(\tau, \sqrt{4\tau}+s\sqrt{\tau}/\tilde L^{2/3})\approx {\rm Ai} \left(-s\right)
\ee



The characteristic determinant (\ref{DETX}) and the  inverse characteristic determinant 
capture the overall depletion of the eigenvalues 
at the edges~\cite{AKE,MU,FIODOROV}. This depletion is universal and for $\beta_D=2$
it also follows from  the method of orthogonal polynomials. Either way, the result is~\cite{FIODOROV,REF202FORR}

\bea
\label{EDGEE}
&&\frac 1{\tilde L^{\frac 23}}\rho(E=\sqrt{4\tau}+s\sqrt{\tau}/\tilde L^{\frac{2}{3}})\approx\nonumber\\
&&
\frac {\tilde L}{\sqrt{\tau}}\left(\left({\rm Ai}^{\prime}(-s)\right)^2+s\left({\rm Ai}(-s)\right)^2\right)
\equiv \frac {\tilde L}{\sqrt{\tau}}f(s)\nonumber\\
\eea
The universal contribution of (\ref{EDGEE}) to the partition function at low temperature is 
($\lambda_0=\sqrt{4\tau}$)

\be
Z[\beta, L,\tau]\approx  \tilde L\,e^{-\beta \lambda_0}
\int_{0}^{\infty }ds\, e^{-\beta \sqrt{\tau}s/\tilde L^{\frac{2}{3}}}\,f(s)
\ee
For large $L$,  the integration is  dominated by the large $s$-asymptotic of theAiry functions

\be
f(s)\approx\sqrt{s}-\frac{1}{4s}\cos\left(\frac{4s^{\frac{3}{2}}}{3}\right)
\ee
The first term yields the leading contribution to the partition function

\be
Z[\beta, L,\tau]\approx \frac{\sqrt\pi}{2}\frac{\tilde L^2}{(\beta\sqrt{\tau})^{\frac 32}}e^{-\beta\lambda_0}
\ee
which results in a leading contribution to the entropy at low temperature

\be
\label{SHIGH}
S\approx N{\rm ln}2 -\frac 32{\rm ln}\left({\beta\lambda_0}\right)
\ee
This is twice the entropy noted in the holographic regime in leading order. In the random matrix regime  ($q^2/N\gg 1$)  the number
of random degrees of freedom grows as $L^2$ and not as $L$.  Finally,  we observe that in the holographic regime ($q^2/N\ll 1$),
the analogue of (\ref{SHIGH}) in the large $N$ limit is

\be
\label{SLOW}
S\approx 2\pi E_N-\frac 32 {\rm ln}(\beta E_N)
\ee
with $E_N/N=e_0\approx 0.0406$~\cite{MALDACENA,POLCHINSKY}. We note that 
(\ref{SLOW}) is of the form suggested  in the context of  the correspondence between a 
black hole  and a highly excited string, with $E_N$ identified as the Rindler energy~\cite{SUSSKIND}. 
In contrast, (\ref{SHIGH}) is dual to a black hole only if the (negative) ground state energy 
or spectrum edge $\lambda_0\rightarrow N{\rm ln}2/(2\pi)$ for $q^2/N\gg 1$. This can  be checked numerically.

\subsection{Spectral and thermal relaxations}

In the random matrix regime, all states are chaotic in leading order, 
and we may ask about their stochastic relaxation
(analogue of quasi-normal modes). For that we note that the correspondence with random matrix theory
allows us to map the SYK evolution of eigenvalues to that of a fluid of eigenvalues~\cite{HYDRO}.
The fluid deformation and relaxation are controlled by the local conservation of the 
density of  eigenvalues and Eulerian dynamics in 1+1 dimension. 
In particular, the local  spectral speed of sound can be read from~\cite{HYDRO} as
$c_s\approx {4\beta_D}/{\sqrt{4\tau}}$ 
with $\beta_D=1,2,4$ (Dyson index).
The characteristic  relaxation time  is the time it takes the sound density 
wave to cross the semi-circle,

\be
T_R\approx \frac {\sqrt{4\tau}}{c_s}=\frac{\tau}{\beta_D}
\ee

Finally, in the holographic regime, the observation that (\ref{SLOW}) is  the entropy of a highly excited string on the
Rindler horizon, suggests 
that the approach to thermal equilibrium in the SYK model is captured in the dual picture by the analogue of an in-falling
string  on a thermal black-hole~\cite{SUSS,QIAN}. For the latter, the entropy grows with the longitudinal momentum 
of the in-falling matter 
at the Rindler horizon $S(t)\approx P(t)\approx e^{\lambda_L t}$ with $\lambda_L=2\pi/\beta$~\cite{SUSS,MALDA}. 
In general, we conclude that the increase in the rate of the logarithm of the entropy
is bounded by  the Kolmogorov-Sinai (rate) entropy, i.e. ${d{\rm ln}S}/{d t}\leq \lambda_L$.
This is the chaos bound reported in~\cite{MALDA} which is reminiscent of the Bekenstein bound~\cite{BEKENSTEIN}.
Both saturates near a black hole. In contrast, a much smaller entropy rate was noted
 in classical chaotic systems far from equilibrium~\cite{BARA}.  
For the SYK model in the holographic regime, the time for the loss of quantum coherence (scrambling time)
$t_L\approx 1/\lambda_L\approx L^0$ is at the edge of the  ballistic regime. It is comparable to the thermalization times 
reported in the  (higher dimensional) holographic models~\cite{EDHOLO,JANIK,MORE}.

\section{Conclusions}

In the holographic regime with $q^2/N\ll 1$,
we have presented qualitative arguments in support of the mesoscopic nature of the SYK model. 
In particular, the pre-chaotic phase appears  super-diffusive.
In the opposite regime with $q^2/N\gg 1$ the SYK model maps on random matrix theory in leading order.
We have shown that in the ergodic phase, the characteristic determinant obeys a viscid Burgers equation
with a small spectral viscosity $\nu_L=1/L$. In this regime, all the SYK spectrum is chaotic 
 with universal Airy oscillations at the edges, in leading order. The  characteristic spectral relaxation of the low lying modes 
is controlled by the spectral speed of sound.

While all our analysis was carried out for the 
complex representations with $\beta_D=2$ universality, we expect that
the results carry for $\beta_D=1,4$ including the soft edge universality,
after careful analysis of the real and quaternion
representations of $Cl([\frac N2])$. An open problem is how the planar approximation 
established in the random matrix regime, can be used to organize 
the quantum mechanical.
Finally,  in the holographic regime, the analogy 
between  the emergent black hole and  a mesoscopic "quantum dot"  offers the 
intriguing possibility for its realization  in other mesocopic systems. The model  may provide
interesting insights for the estimate of  thermalization times in current collider experiments.

\section{Acknowledgements}

This work was supported by the U.S. Department of Energy under Contract No.
DE-FG-88ER40388 (YL,IZ) and by the Grant DEC-2011/02/A/ST1/00119 of the Polish National Center of Science (MAN).

 \vfil

\end{document}